\documentclass[11pt]{article}
\usepackage{amsmath}
\usepackage{epsfig}
\usepackage{amssymb}
\usepackage{graphicx}
\usepackage{algorithmic}
\usepackage{subfigure}
\usepackage{url}
\usepackage{wrapfig}
\usepackage[ruled,vlined,linesnumbered]{algorithm2e}
\usepackage{color}

\makeatletter
\renewcommand{\paragraph}{\@startsection{paragraph}{4}{0ex}%
   {-3.25ex plus -1ex minus -0.2ex}%
   {1.5ex plus 0.2ex}%
   {\normalfont\footnotesize\bfseries}}
\makeatother

\begin{document}

\mathchardef\Gamma="0100
\mathchardef\Delta="0101
\mathchardef\Theta="0102
\mathchardef\Lambda="0103
\mathchardef\Xi="0104
\mathchardef\Pi="0105
\mathchardef\Sigma="0106
\mathchardef\Upsilon="0107
\mathchardef\Phi="0108
\mathchardef\Psi="0109
\mathchardef\Omega="010A

\newcommand{\user}{\mbox{$u$}}
\newcommand{\ulevel}{\mbox{$\bar{u}$}}
\newcommand{\bigsqcap}{\mbox{{\Large $\sqcap$}}}
\newcommand{\dl}{\mbox{$\, [ \hspace*{-1.5pt} [\,$}}
\newcommand{\dr}{\mbox{$\, ] \hspace*{-1.5pt} ]\:$}}
\newcommand{\da}{\mbox{$\, A \hspace*{-6.75pt} A \,$}}
\newcommand{\drightarrow}{\mbox{$\rightarrow \hspace*{-8pt} 
\rightarrow$}}

\newcommand{\umodels}{\mbox{$\models_{\ulevel}$}}
\newcommand{\hmodels}{\mbox{$\models_{{\bf H},\bar{u}}$}}
\newcommand{\imodels}{\mbox{$\models_{I,\bar{u}}$}}
\newcommand{\tmodels}{\mbox{\,$\models_{{\bf H}}$\,}}
\newcommand{\db}{\mbox{\,$\langle \Delta, \bar{u}\rangle$\,}}
\newcommand{\tp}{\mbox{\,${\bf T}_{\Delta}^{\bar{u}}$\,}}
\newcommand{\tpi}{\mbox{\,$T_\Delta^{\bar{u}}$\,}}
\newcommand{\md}{\mbox{\,${\bf M}_{\Delta}$\,}}

\newtheorem{defn}{Definition}[section]
\newtheorem{thrm}{Theorem}[section]
\newtheorem{prop}{Proposition}[section]
\newtheorem{lemm}{Lemma}[section]
\newtheorem{obsv}{Observation}[section]
\newtheorem{corr}{Corollary}[section]
\newtheorem{example}{Example}[section]

\newcommand{\vs}{\vspace{1ex}}              
\newcommand{\vsp}{\vspace{2ex}}             
\newcommand{\vspp}{\vspace{4ex}}            

\newcommand{\negvs}{\vspace*{-1ex}}         
\newcommand{\negvsp}{\vspace*{-2ex}}        
\newcommand{\negvspp}{\vspace*{-4ex}}       

\newcommand{\hs}{\hspace*{1em}}                      
\newcommand{\hsp}{\hspace*{2em}}                     
\newcommand{\hspp}{\hspace*{4em}}                    

\newcommand{\ie}{\mbox {\em i.e.}}
\newcommand{\eg}{\mbox {\em e.g.}}
\newcommand{\Eg}{\mbox {\em E.g.}}

\def\blackbox{{\rule{2.2mm}{2.2mm}}}
\def\qed{\hspace*{\fill}\blackbox}

\newcommand{\definitionbox}{\qed}
\newcommand{\examplebox}{\qed}
\newcommand{\lemmabox}{\qed}
\newcommand{\theorembox}{\qed}
\newcommand{\conjecturebox}{\qed}
\newcommand{\algorithmbox}{\qed}
\newcommand{\observationbox}{\qed}
\newcommand{\obsbox}{\qed}
\newcommand{\assumptionbox}{\qed}
\newcommand{\notesbox}{\qed}

\newcommand{\ra}{\mbox{$\rightarrow$}}
\newcommand{\la}{\mbox{$\leftarrow$}}

\newcommand{\fd}{\mbox{$\rightarrow$}}

\newcommand{\munion}{\mbox{$\cup^m$}}
\newcommand{\join}{\mbox{$\bowtie$}}
\newcommand{\distinct}{\mbox{{\tt distinct}}}

\newcommand{\ojoin}{\mbox{${\Join}^o$}}

\newcommand{\xssql}{\mbox{{\bf Xssql}}}

\newcommand{\BA}{\mbox{$\textit{\textbf{a}}$}}
\newcommand{\BB}{\mbox{$\textit{\textbf{b}}$}}
\newcommand{\BC}{\mbox{$\textit{\textbf{c}}$}}
\newcommand{\BD}{\mbox{$\textit{\textbf{d}}$}}
\newcommand{\BE}{\mbox{$\textit{\textbf{e}}$}}
\newcommand{\BF}{\mbox{$\textit{\textbf{f}}$}}
\newcommand{\BO}{\mbox{$\textit{\textbf{o}}$}}
\newcommand{\BP}{\mbox{$\textit{\textbf{p}}$}}
\newcommand{\BQ}{\mbox{$\textit{\textbf{q}}$}}
\newcommand{\BR}{\mbox{$\textit{\textbf{r}}$}}
\newcommand{\BV}{\mbox{$\textit{\textbf{v}}$}}
\newcommand{\BL}{\mbox{$\textit{\textbf{l}}$}}
\newcommand{\BI}{\mbox{$\textit{\textbf{i}}$}}
\newcommand{\BH}{\mbox{$\textit{\textbf{h}}$}}
\newcommand{\BS}{\mbox{$\textit{\textbf{s}}$}}
\newcommand{\BK}{\mbox{$\textit{\textbf{k}}$}}
\newcommand{\BT}{\mbox{$\textit{\textbf{t}}$}}
\newcommand{\BX}{\mbox{$\textit{\textbf{x}}$}}

\begin{center}
{\Large {\bf Reliable Querying of Very Large, Fast Moving and Noisy Predicted Interaction Data using Hierarchical Crowd Curation}}\\
{\bf Hasan Jamil}$^\natural$ and {\bf Fereidoon Sadri}$^\flat$\\
$^\natural$Department of Computer Science, University of Idaho, USA\\
{\tt jamil@uidaho.edu}\\
$^\flat$Department of Computer Science, University of North Carolina at Greensboro, USA\\
{\tt f\_sadri@uncg.edu}
\end{center}

\begin{abstract}
The abundance of predicted and mined but uncertain
biological data show huge needs for massive, efficient
and scalable curation efforts. The human expertise
warranted by any successful curation enterprize is often
economically prohibitive especially for speculative end
user queries that may not ultimately bear fruit. So
the challenge remains in devising a low cost engine
capable of delivering fast but tentative annotation and
curation of a set of data items that can be authoritatively
validated by experts later demanding significantly small
investment. The aim thus is to make a large volume of
predicted data available for use as early as possible
with an acceptable degree of confidence in their accuracy
while the curation continues. In this paper, we present
a novel approach to annotation and curation of biological
database contents using crowd computing. The technical
contribution is in the identification and management of
trust of mechanical turks, and support for ad hoc declarative
queries, both of which are leveraged to support reliable
analytics using noisy predicted interactions.
\end{abstract}

%
%


\section{Introduction}
\label{introduction}

Protein-protein interactions (PPI) are one of the central biological activities in almost all cellular processes. Thus, accurate structural and functional annotations of PPI are vital to understand their biological and pathological implications for new therapeutics design \cite{MoalJF15}. PPI network analyses are also often used to computationally predict protein functions to decipher molecular mechanisms and biological processes as an inexpensive alternative to more expensive laboratory methods \cite{Peng2014}. PPI networks essentially provide a systems-level understanding of the underlying disease mechanisms, metabolic functions, help identify novel bio-markers, etc. \cite{Sevimoglu201422}. Collection, curation and maintenance of such networks thus play a critical role in biological data analysis in today's post-genomic era.

\begin{figure*}[ht]
\centering
\subfigure[Actions.]
{\label{act}
\includegraphics[height=1.5in,width=.31\textwidth]{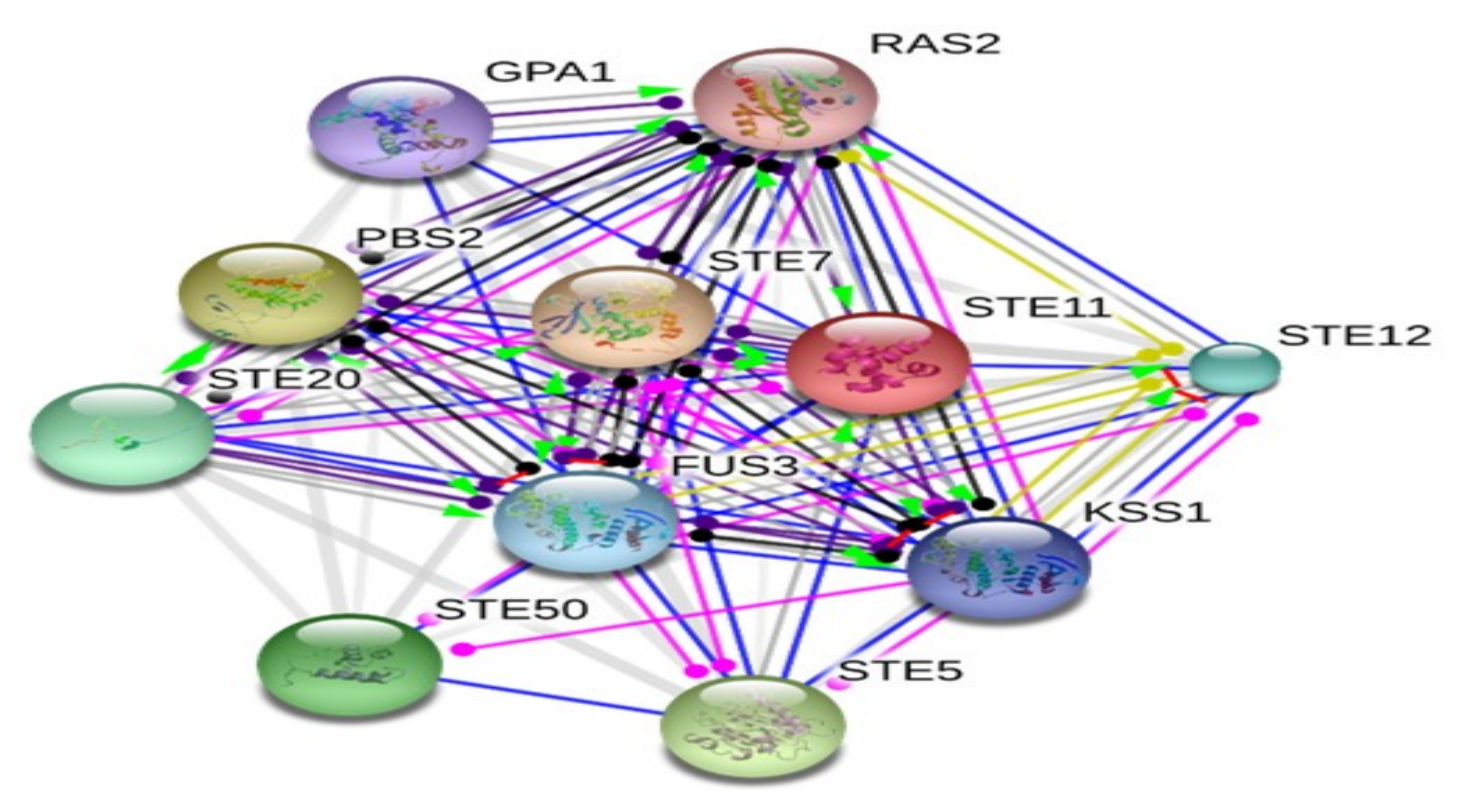}
}
\subfigure[Evidence.]
{\label{evi}
\includegraphics[height=1.5in,width=.31\textwidth]{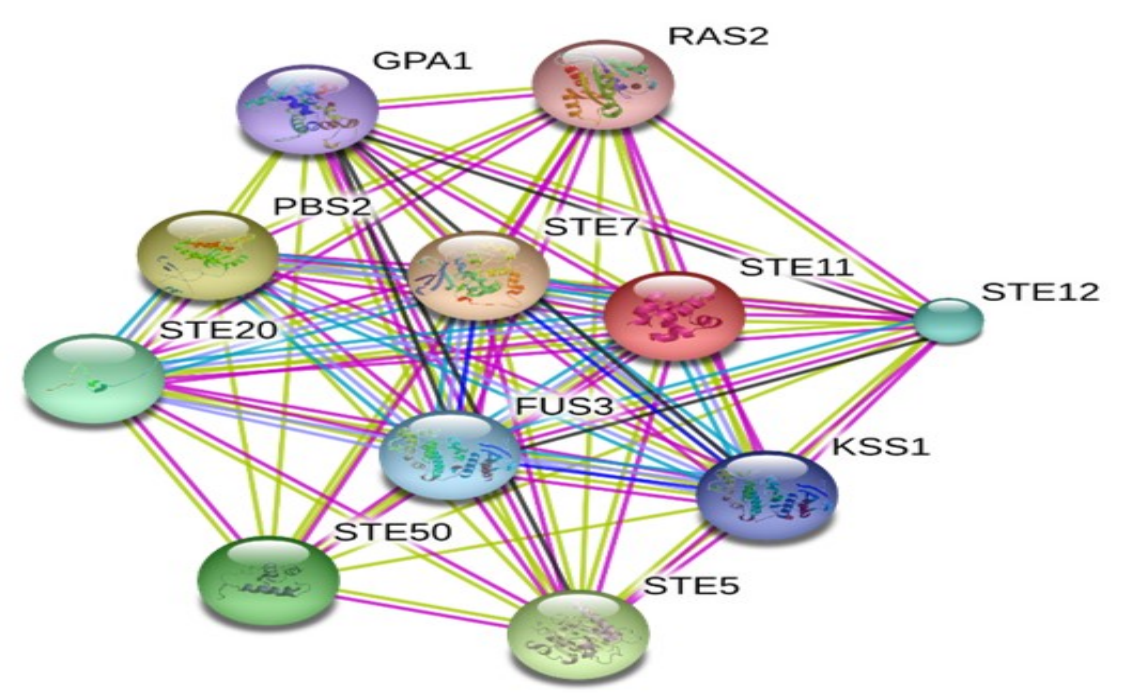}
}
\subfigure[Confidence.]
{\label{con}
\includegraphics[height=1.5in,width=.31\textwidth]{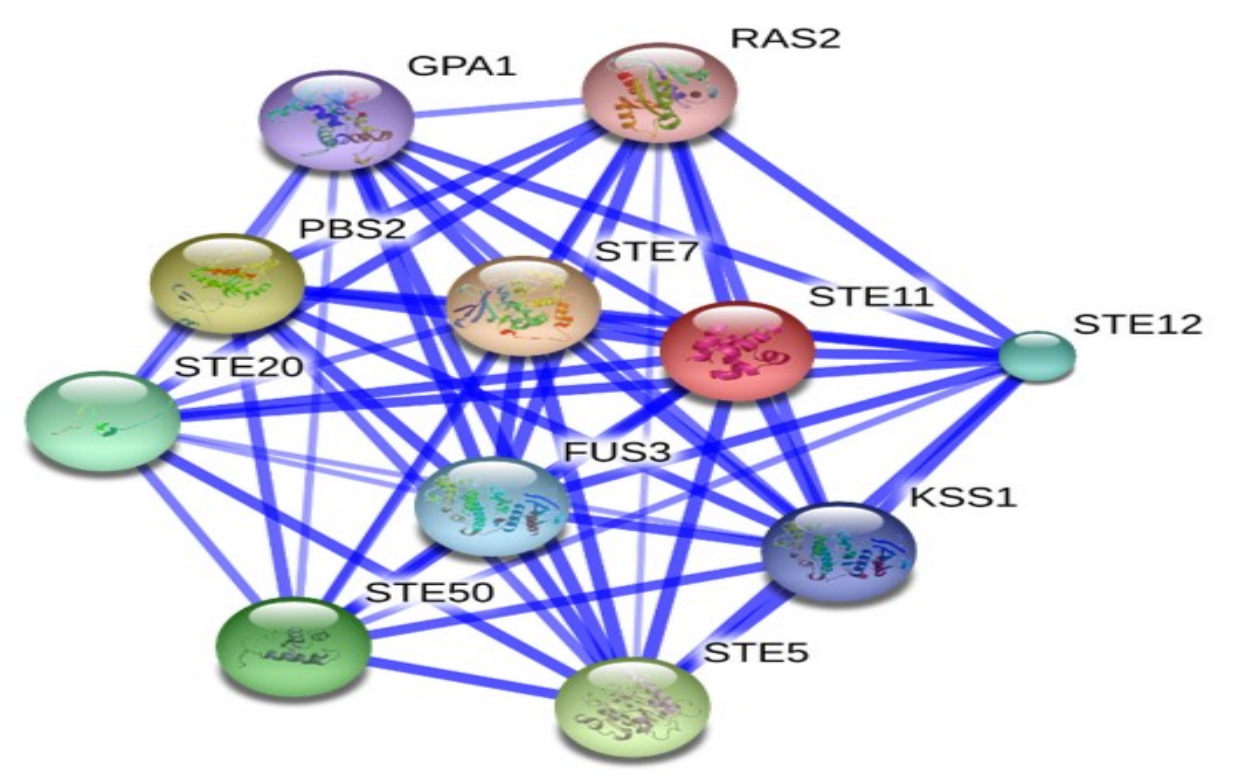}
}
\subfigure[Overall STE11 prediction information for top ten interactors.]
{\label{over}
\centerline{\fbox{\includegraphics[height=2.25in,width=.93\textwidth]{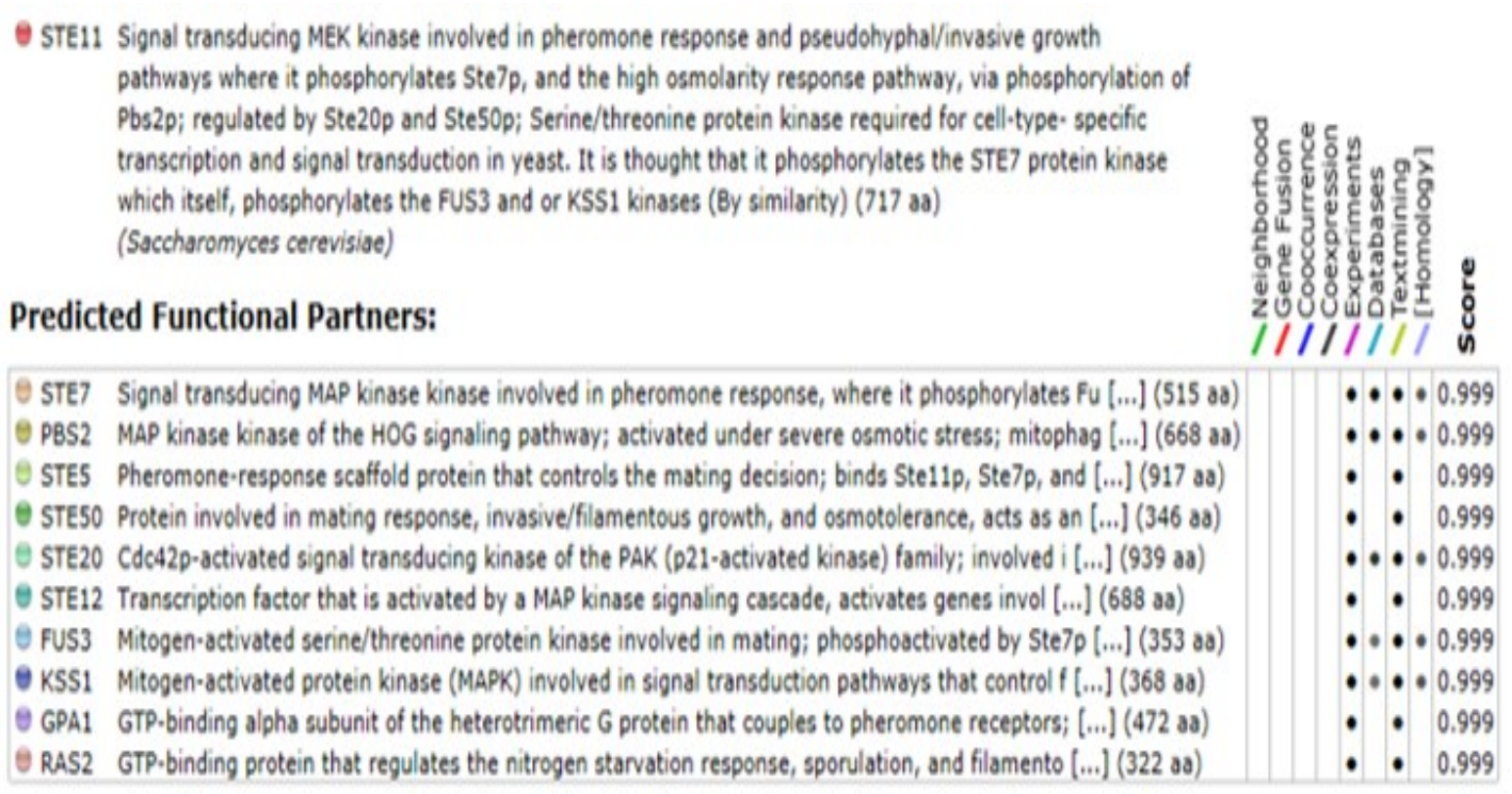}}}
}
\caption{Various interaction graph views of protein STE11.} \label{STE11-int}
\end{figure*}

To this end, researchers have been collecting interaction data to store in databases and annotate them for various applications for sometime \cite{Szklarczyk2015,Orchardetal14}. Integrating predicted and less accurate interaction data with more reliable laboratory tested data has been an approach essentially to increase the coverage of the network analyses for novel discoveries \cite{ZhangLYW15}. To expedite discovery, researchers have been increasingly adopting more predicted interactions \cite{SubramaniKMN15} into their analyses obtained from high throughput data generation systems such as next generation sequencing \cite{citeulike:13903650}, yeast two-hybrid screens \cite{citeulike:789740,Mehla01052015} mass spectrometry \cite{citeulike:789741} and literature mining \cite{HaiderLPLDKLG15}. Unfortunately the high error rates in these predicted data sets invite two major challenges. First, use of these datasets render the analysis highly speculative and often wrong. Secondly, until these datasets are curated \cite{CaoXXC15} beyond an acceptable level of certainty, they essentially remain less than useful.

Manual or machine curation of these predicted interactions is often a monumental task. Additionally, autonomous \cite{AntonyBHP08}, semi-autonomous \cite{JamiesonGSNR12} and manual curation \cite{KwonKSCW14} have varying degrees of accuracy rates along with their respective scale factors. The BioCreative II effort \cite{citeulike:3194296} recently studied the difficulties associated with the curation and annotation process of PPI data. They found that only about 35\% precision and recall could be achieved for mapping interactions obtained from text mining to SwissProt, and only about 20\% of predicted interactions from full-text article mining could be matched with human curator identified sentences. Such low success rates reflect the difficulty of obtaining meaningful and accurate curation, and the scopes that exist to design novel ways to curate or use the uncurated interactions. The need is so pressing that it motivated the desperate attempt to even manually curate about 88,000 documents \cite{DavisWRKLLSJKGHMEM13}. In this paper, our goal is addressing this niche and propose a method to make uncurated interaction data meaningfully usable in an evolving database.

\subsection{An Illustrative Example using STRING and eIST}

Consider extracting all interactors of the Serine/threonine-protein kinase STE11 in Saccharomyces cerevisiae from the interaction database STRING \cite{Szklarczyk2015}. STRING supports multiple types of queries to search all known and predicted interactors of a given set of proteins or gene products. Figures \ref{act} through \ref{con} show three views of ten most prominent interactors of the STE11 in a single protein query mode and their summary characterizations are shown in figure \ref{over}.

\begin{figure*}
\centerline{\includegraphics[height=2.25in,width=.93\textwidth]{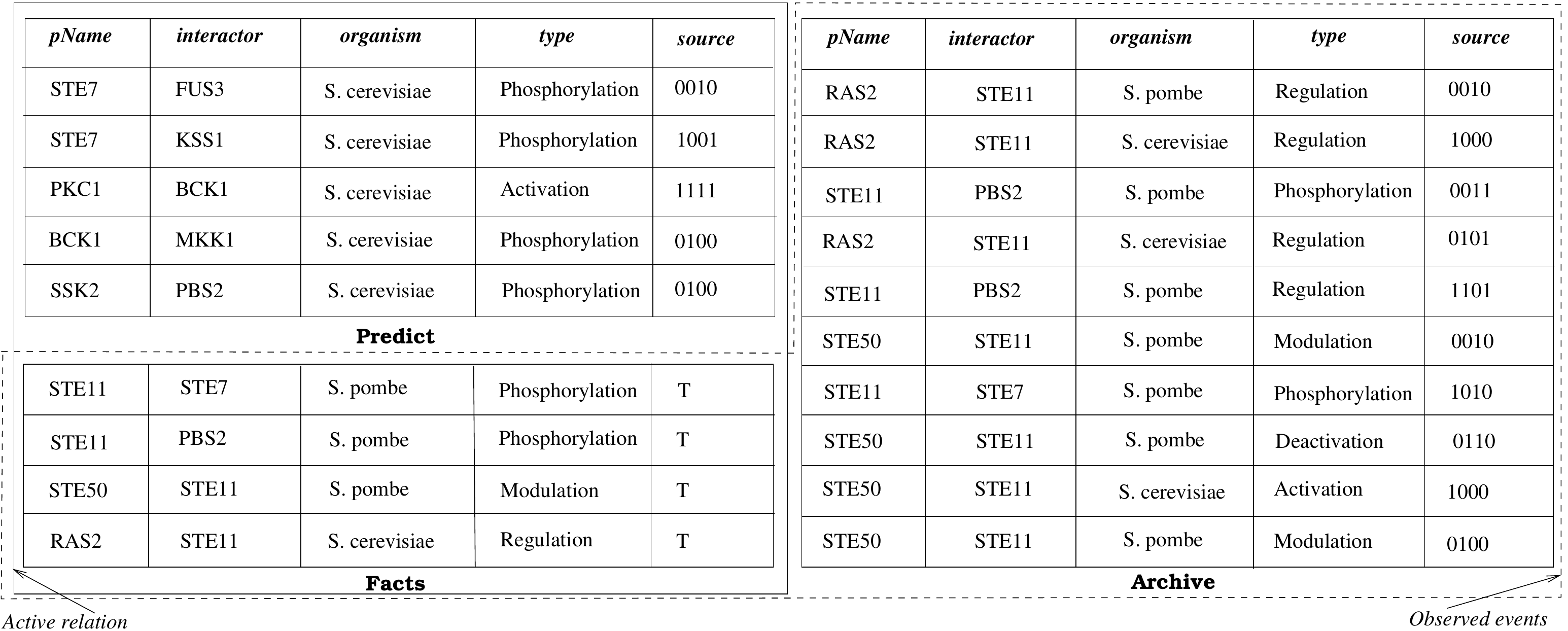}}
\caption{An eIST interaction database {\bf D}.} \label{nets}
\end{figure*}

We can imagine that the three views of the interacting partners of STE11 in figures \ref{act} through \ref{con} are likely generated from a traditional relational tabular view of the interactions functionally similar to the extended information source tracking (eIST) \cite{UncMan-ismis-1994} tables as shown in figure \ref{nets}. In this earlier research on eIST, all relations are composed of three complementary tables having identical schema -- {\em Predict, Facts}, and {\em Archive}. The tuples in {\em Facts} are regular relational tuples, except that all eIST tables have an additional system recognized attribute called {\em source} which contains a vector of sources contributing the tuple.

Assuming that we have four contributing curators, or sources, in this example database -- {\em Russ, Fred, Karen} and {\em Maria} in this order, and of course the special system source {\tt T}, the tuple {\tt <STE11,STE7,S. pombe,Phosphorylation@1010>} in {\em Archive} means that {\em Russ} and {\em Karen} contributed this tuple, and the validity and reliability of it now depends on the combined reliability of both contributors. However, its presence in {\em Archive} signifies that it is an {\em image} of a {\tt true} tuple in {\em Facts} and thus does not take part in any query. Tuples in {\em Archive} migrate from {\em Predict} every time a {\tt true} tuple corresponding to it is added in the {\em Facts} table, and are only used to determine the credibility of the sources based on their prediction or curation performance. For example, if we add {\tt <BCK1,MKK1, S.~cerevisiae,Activation@T>} in {\em Facts}, the tuple {\tt <BCK1,MKK1,S.~cerevisiae, Phosphorylation@0100>} in {\em Predict} will migrate to {\em Archive} since the primary key components of both tuples (i.e., {\tt <BCK1,MKK1, S.~cerevisiae>} agree, and once we know that {\tt <BCK1,MKK1, S.~cerevisiae,Activation>} is a fact (i.e., 100\% reliable), there is no reason for {\tt <BCK1,MKK1,S.~cerevisiae, Phosphorylation @0100>} to continue to exist in {\em Predict}. The source vector {\tt T} is a special system source which is always correct, and thus is the $n+1$th source. Therefore, for the tuple {\tt <BCK1,MKK1,S.~cerevisiae,Phosphorylation @0100>}, vector {\tt <0100>} in reality is {\tt <01000>} and for tuple {\tt <BCK1,MKK1, S.~cerevisiae, Activation@T>}, it is {\tt <00001>}, with the proviso that whenever $n+1$ bit of a vector is 1, the rest must be zeros, it is 0 otherwise.

The basic IST model was introduced earlier in \cite{Sadri91-tkde}, and the algebraic model and reliability calculation method of eIST has been later devised by Jamil and Sadri in \cite{UncMan-ismis-1994} to accommodate the time varying nature of source reliability based on past performance. In these two works, it had been shown that these models are capable of associating a confidence score to any query tuple computed using arbitrary {\tt Select Project Join} (SPJ) query based on eIST relations. The eIST model includes source vector manipulations for all basic relational algebra operations, and a reliability calculation algorithm based on source credibility. These models provide the ability to determine the {\em provenance} of each query tuple. That is, how the tuple was obtained and which sources contributed. In itself, it is a valuable tool for explaining how the answer was derived. In addition, a probability or certainty factor can be computed for each query tuple using the provenance. IST and eIST use solid mathematical probability theory for the probability calculation. But it is easy to accommodate other approaches \cite{LakshmananS01b}, such as the fuzzy logic technique, for the computation of certainty factors. We omit a detailed discussion of these models here for brevity and refer the readers to \cite{UncMan-ismis-1994,Sadri91-tkde} for an in-depth discussion.

\begin{example}[Calculating Tuple Reliability]
Consider the tuple {\tt <BCK1, MKK1, S.~cerevisiae, Phosphorylation@0100>} in {\em Predict}, which shows that source 2, {\em Fred}, contributed to this information. The reliability of sources are maintained and updated periodically in eIST. Assume the reliability of {\em Fred} has been determined to be 0.8. Then the certainty factor of the tuple {\tt <BCK1, MKK1, S.~cerevisiae, Phosphorylation@0100>} is 0.8, or 80\%. Now assume that this information was contributed by {\em Fred} and by {\em Karen}, which would be represented by tuple {\tt <BCK1,MKK1, S.~cerevisiae, Phosphorylation@0110>} in {\em Predict}. If reliability of {\em Karen} has been determined to be 0.85, we expect the certainty of this tuple to be very high, since these sources are {\em independently} confirming this information. The certainty factor of {\tt <BCK1,MKK1, S.~cerevisiae, Phosphorylation@0110>} is computed as
$1 - (1-0.85) \times (1 - 0.8) = 0.97$ in the probabilistic approach. Note that in the fuzzy logic approach, the certainty factor of this tuple is computed as the higher reliability of contributing sources, namely, 0.85. Fuzzy logic formulation may be relevant in some applications, and eIST can accommodate either approach depending on users preference.
\end{example}

However, the observation we would like to make here is that the three views, i.e., actions, evidence and confidence, that STRING generates can be easily constructed from an eIST database with the added benefit of a time varying confidence score in a dynamic fashion. The partitioning of the tuples in {\em Predict} and {\em Facts} also makes it possible to start using raw data without serious expert curation as soon as they become available in {\em Predict} and need not be in {\em Facts}. The eight evidential attributes in figure \ref{evi} STRING uses to assign a confidence score in figure \ref{con} can each be modeled as a source and their combined reliability can be used to generate a network similar to figure \ref{con} in eIST. We leverage this observation in this paper to propose a new crowdsourced curation model for PPI networks using a declarative query language, called {\em CureQL}.

\subsection{Using CrowdDB for Content Gathering from People}

Leveraging crowd participation for content gathering is the principal focus of databases such as CrowdDB \cite{FranklinKKRX11} and Qurk \cite{MarcusWMM11}. They support a declarative language to initiate crowd participation for specific data items or human observations such as subjective comparison of images. They also support several ancillary  functions to manage the collection process at the systems level including language support for managing Human Intelligent Task (HIT) and HIT assignments, active time of the crowd queries, and query specific fabrication of the query interfaces. We adapt CrowdSQL (see \cite{FranklinKKRX11} for more details) supported in CrowdDB as a substrate for CureQL to inherit its inherent simplicity and leverage its powerful features.

\begin{figure*}
\centering{
\begin{tabular}{ll}
\begin{tabular}{l}
$c_1$: \texttt{CREATE CROWD TABLE} {\em Interaction} (\\
\hspace*{4mm} \hspace*{3mm} {\em pName} \texttt{STRING},\\
\hspace*{4mm} \hspace*{3mm} {\em interactor} \texttt{STRING},\\
\hspace*{4mm} \hspace*{3mm} {\em organism} \texttt{STRING},\\
\hspace*{4mm} \hspace*{3mm} {\em type} \texttt{STRING},\\
\hspace*{4mm} \hspace*{3mm} \texttt{PRIMARY KEY} ({\em pName, interactor, organism}),\\
\hspace*{4mm} \hspace*{3mm} \texttt{FOREIGN KEY} ({\em pName, interactor, organism}), \\
\hspace*{4mm} \hspace*{3mm} \hspace*{3mm} REF {\em Names} ({\em name, name, organism}) );
\end{tabular} &
\begin{tabular}{l}
$c_2$: \texttt{CREATE TABLE} {\em Names} (\\
\hspace*{4mm} \hspace*{3mm} {\em name} \texttt{CROWD STRING},\\
\hspace*{4mm} \hspace*{3mm} {\em organism} \texttt{STRING},\\
\hspace*{4mm} \hspace*{3mm} {\em date} \texttt{CROWD DATE},\\
\hspace*{4mm} \hspace*{3mm} \texttt{PRIMARY KEY} ({\em name,}\\
\hspace*{4mm} \hspace*{3mm} \hspace*{3mm}  {\em organism}) );
\end{tabular}
\\ \\
\begin{tabular}{l}
$q_1$: \texttt{SELECT} {\em pName, organism}\\
\hspace*{4mm} \texttt{FROM} {\em Interaction}\\
\hspace*{4mm} \texttt{WHERE} {\em type = "activation"};
\end{tabular} &
\begin{tabular}{l}
$q_2$: \texttt{SELECT} {\em name, organism}\\
\hspace*{4mm} \texttt{FROM} {\em Names}\\
\hspace*{4mm} \texttt{WHERE} {\em date $\geq$ "1/31/2016"};
\end{tabular}
\end{tabular}
}
\caption{Crowd specification of database {\bf D} in figure \ref{nets} in CrowdSQL/CureQL.}
\label{exmp1}
\end{figure*}

In the examples in figure \ref{exmp1}, the table in $c_1$ is called a {\em crowd table}, and the table in $c_2$ is a table with {\em crowd columns}, but both are called {\em crowd enabled tables}.
The semantics of the CrowdSQL's \cite{FranklinKKRX11} crowd enabled tables are straightforward -- when a crowd enabled relation is referenced in an query, entire rows are extracted with crowd participation corresponding to a crowd table as opposed to columns extracted corresponding to a table with crowd columns. For example, the query $q_1$ in figure \ref{exmp1} asks the crowd to supply entire rows and add to the existing {\em Interaction} table since the entire table is crowd enabled. However, the query $q_2$ asks for only the the missing values (identified with the presence of a {\tt CNULL}, or crowd NULL value) for the columns {\em name} and {\em date}. Both queries, however, will finally select only the tuples that satisfy the corresponding selection conditions.

\subsection{CrowdCure Approach and Contributions}

While the CrowdDB and eIST frameworks individually support some of the complementary features we envision in a community curation system, several additional features are still necessary to develop an end to end working system, we call {\em CrowdCure}. For example, support for the use of specific extraction tools on target data sources, assignment of desired and relevant curators to sets of predicted data, organizing the curators in an expertise hierarchy, etc. within a declarative setting to aid ad hoc queries.

From an end user standpoint, we expect a user to be able to pose both ad hoc traditional eIST, and crowd curation queries in CrowdCure. By that we mean, when new contributions from crowd aren't necessary, we should be able to pose simple queries and get traditional responses with tuple reliability. Recall that all tables in eIST are potentially uncertain (their certainty ranges between the interval (0,1]) and their certainty is determined by the reliability of the contributing sources, and that all tables have a special attribute called {\em source} and all tables have three possibly empty components {\em Facts, Predict} and {\em Archive}.

We, therefore, adopt the provision of explicit invocation of crowd curation features even on crowd enabled tables. We believe this approach is pragmatic and ensures flexible use of the underlying database to suit the application semantics as needed.
Thus from a systems standpoint, we support an interface in which users can issue any mix of extraction, curation and analysis queries directly using a declarative language. In extraction queries, linguistic features are supported to specify mining tools, set of documents, and to partition extraction tasks for groups of curators organized in a hierarchy. Visual interfaces are supported for each of these tasks. For analysis queries, the traditional features of eIST becomes handy. The approach also supports separation of extraction from a hierarchical curation, and as soon as the data are available for curation, they can be used in queries. As the curation progresses, their reliability improves (positively or negatively) and are reflected in the composite SPJ query response in a dynamic temporal fashion. Finally, another desirable feature of CrowdCure is the ability of eIST to display the current reliabilities it has determined for all sources. We allow users to override eIST reliabilities and supply their own reliabilities for one or more sources if they wish. We also support, among other features, ``what-if" queries allowing users to experiment with the effects of changing source reliabilities.

\section{Related Research}

In CrowdCure, we are dealing with research challenges in two orthogonal axes -- crowd curation, and assignment of reliability to curated interaction data. A related issue is how to quantify trust of computed results of arbitrary queries using the base inaccurate data curated by numerous potentially unreliable experts. The challenge is to marry these two orthogonal advances into one single coherent platform for reliable crowd curation and supporting queries with quantifiable reliability in a user transparent manner.

Collaborative or crowd computing rely on social participation of masses toward a common goal such as content curation. Two most widely employed curation types are expert-based annotation and machine generated classification. Traditionally, however, the curation was solely an expert's job. Algorithmic curation, on the other hand, ranks curation items such as relevant web pages for search queries or produce recommendations on e-commerce databases for user consumptions. In contrast the voting based annotation commonly employed in e-commerce applications is high in popularity due to its simplicity and amenability to natural ranking \cite{AskalidisS2013} because the upvote and downvote based ranking of curation overcomes a few drawbacks of expert-based curation (cost of experts and scale) and algorithm based curation (inflexibility of algorithms of content types). However, the voting based curation has not been shown to be effective in technical content ranking which demand their understanding scientifically. Technically, interaction annotation require human understanding of an interaction pair in a document that is potentially confusing and fundamentally distinct from an overall likability of an entire document or a piece of item. In other words, technical curation usually involves extraction of structured data with various attributes from texts to be stored in a structured database such as relational database.

\begin{figure*}[ht]
\centering{
\begin{tabular}{ll}
\begin{tabular}{l}
$q_3$: \texttt{SELECT} {\em pName, interactor, type}\\
\hspace*{4mm} {\tt USING} {\em UniHi} \texttt{ON} {\em I\_papers}\\
\hspace*{4mm} \texttt{SOURCE} $r_1$ \texttt{BEFORE} \texttt{SELECT} {\em eName}\\
\hspace*{4mm} \hspace*{3mm} \hspace*{3mm} \hspace*{3mm} \hspace*{3mm} \texttt{FROM} {\em Experts}\\
\hspace*{4mm} \hspace*{3mm} \hspace*{3mm} \hspace*{3mm} \hspace*{3mm} \texttt{WHERE} {\em laboratory = "MIT"};\\
\hspace*{4mm} \texttt{FROM} {\em Interaction}\\
\hspace*{4mm} \texttt{WHERE} {\em type = "activation"};
\end{tabular} &
\begin{tabular}{l}
$c_3$: \texttt{CREATE TABLE} {\em Experts} (\\
\hspace*{4mm} \hspace*{3mm} {\em eName} \texttt{STRING},\\
\hspace*{4mm} \hspace*{3mm} {\em species} \texttt{STRING},\\
\hspace*{4mm} \hspace*{3mm} {\em laboratory} \texttt{STRING},\\
\hspace*{4mm} \hspace*{3mm} {\em eMail} \texttt{STRING UNIQUE},\\
\hspace*{4mm} \hspace*{3mm} \texttt{SOURCE KEY} ({\em eName}), \\
\hspace*{4mm} \hspace*{3mm} \texttt{PRIMARY KEY} ({\em eName, species}) );
\end{tabular}
\\ \\
\begin{tabular}{l}
$q_4$: \texttt{SELECT} {\em t.pName, t.interactor, t.type}\\
\hspace*{4mm} {\tt USING} {\em UniHi} \texttt{ON} {\em I\_papers}\\
\hspace*{4mm} {\tt LIMIT TIME} {\em t} {\tt UNITS}, {\tt DATA} {\em k} {\tt ROWS}\\
\hspace*{4mm} \texttt{FROM} {\em Interaction} {\tt AS} {\em y}\\
\hspace*{4mm} \texttt{GROUP BY} {\em type}\\
\hspace*{4mm} \texttt{CLUSTER SOURCE} (\texttt{SELECT} {\em eName, laboratory}\\
\hspace*{4mm} \hspace*{3mm} \texttt{FROM} {\em Experts}\\
\hspace*{4mm} \hspace*{3mm} {\tt WHERE} {\em y.organism = species}) \texttt{BEFORE} \\
\hspace*{4mm} \hspace*{3mm} (\texttt{SELECT} {\em eName, species}\\
\hspace*{4mm} \hspace*{3mm} \texttt{FROM} {\em Experts});
\end{tabular} &
\begin{tabular}{l}
$c_4$: \texttt{CREATE VIEW} $r_1$ \texttt{AS}\\
\hspace*{4mm} \hspace*{3mm} \texttt{SELECT} {\em eName}\\
\hspace*{4mm} \hspace*{3mm} \texttt{FROM} {\em Experts}\\
\hspace*{4mm} \hspace*{3mm} \texttt{WHERE} {\em laboratory = "Yale"};\\ \\
$q_5$: \texttt{SELECT} {\em name, organism}\\
\hspace*{4mm} {\tt USING} {\em PIPs} \texttt{ON} {\em P\_papers}\\
\hspace*{4mm} \texttt{SOURCE} $r_2$ \texttt{BEFORE} $r_3$\\
\hspace*{4mm} \texttt{FROM} {\em Names};
\end{tabular}
\end{tabular}
}
\caption{CureQL queries over the database {\bf D} (continued from figure \ref{exmp1}).}
\label{exmp2}
\end{figure*}

Research in collecting structured data has gained its due attention recently \cite{ParkW14}. In these research, turks are asked to identify structured data in texts in order to populate structured tables in ways similar to extracting interaction data from free texts \cite{SubramaniKMN15}. The process of extracting interaction data by turks is facilitated by suitable mining and text highlighting interfaces in some research to help extract quality data \cite{RahmanianD14,NakatsuI14}. Several research also tried to quantify the quality or reliability of extracted data in some fashion \cite{AlonsoMN13a}, and several adopted an alternative method of algorithmic validation of extracted data \cite{JamiesonGSNR12,AntonyBHP08}. An entirely different approach avoids assignment of reliability to or validation of extracted data by computationally choosing reliable turks in the first place \cite{BozzonBCSV13}. Our understanding is that none of these approaches scale well for large data sets, exhibit unreliable certainty assignment and skewed quantification \cite{XieHZGFY15}.


\section{CureQL Language and its Semantics}

Technically, CureQL extends CrowdSQL in two ways: as shown in figure \ref{exmp2}, (i) it introduces the notion of {\tt SOURCE KEY} in the {\tt CREATE TABLE} and {\tt CREATE CROWD TABLE} statements to declare a set of attributes to be identifiers a la object-oriented systems to be used as curator identities, and (ii) it introduces an optional {\tt USING ON} clause with an optional {\tt LIMIT} and a mandatory {\tt SOURCE} subclause, with the proviso that the {\tt SOURCE} subclause may be used with the {\tt USING} clause globally, or with the {\tt GROUP BY} clause for every partition locally.

Consider the use of {\tt USING ON} clause and the {\tt SOURCE} subclause in query $q_3$ in figure \ref{exmp2}. {\tt USING} subclause has no effect on a query if it does not include a crowd column or the table is not a crowd table. When used, it expects an analytics, such as {\em UniHi} \cite{Kalathur08112013} (query $q_5$ uses {\em PIPs} \cite{McDowall01012009}, {\em ChiliBot} \cite{Chen2004} is another example), which requires an input {\em I\_papers}. The {\tt ON} modifier is optional to allow analytics without input needs. However, the analytics must return all the crowd columns for the {\tt SELECT} query to be successful. Since {\tt USING} is an extraction operation, and all extracted data possibly are predictions (all crowd contributions are treated as uncertain), all tuples now must be tagged with a set of contributing sources or curators. In CrowdQL, the first set of contributors are also treated as curators. This tagging is accomplished using the mandatory {\tt SOURCE} subclause.

The {\tt SOURCE BEFORE} subclause expects an arbitrary but totally ordered sets of source keys to serve as the contributing or curating experts of the predicted data. Each scheme $R_i$ of relations $r_i$ in the subclause of the form {\tt SOURCE $r_1$ BEFORE $\ldots$ BEFORE $r_k$} must contain a unique {\tt SOURCE KEY}. The projection of the {\tt SOURCE KEY}s are the assigned curators for each extracted tuple in the {\em Predict} set. However, the set of curators in $r_j$ are strictly at a higher level than the curators in set $r_i$ for any $i<j$ and are more credible. As query $q_3$ shows, CureQL supports supplying these sets as either a ground relation, a view (as in the statement $c_4$) or as a CureQL query (in queries $q_3$ and $q_4$) satisfying the {\tt SOURCE KEY} requirement. In query $q_5$, if none of the attributes of $r_2$ (or $r_3$) is part of a {\tt SOURCE KEY}, the entire scheme is treated as one.

The scope rule of {\tt SOURCE BEFORE} subclause can be used in two interesting ways. First, the relations in the {\tt FROM} clause in query $q_3$ do not range over the {\tt SOURCE BEFORE} subclause as it appeared before the {\tt FROM} clause, and thus the use of tuple variables in the CureQL queries in {\tt SOURCE BEFORE} subclause has no effect. Second, in the query $q_4$, the relation {\em Interaction} has scope over the {\tt SOURCE BEFORE} subclause that follows the {\tt FROM} clause and thus influences the selection of the experts in {\tt SOURCE} subclause based on the species value for the first curator set in the order. Thus, the the use of {\tt CLUSTER} modifier before the {\tt SOURCE BEFORE} subclause in {\tt GROUP BY} clause forces assigning potentially different sets of curator hierarchies for every group partitioned by the {\tt GROUP BY} statement. However, note that the {\tt SOURCE} clause cannot be used in both {\tt USING} and {\tt GROUP BY} clauses simultaneously. Finally, in order to deal with uncertainties of long crowd response delay and potentially large volumes, CureQL supports the {\tt LIMIT} option slightly differently than Qurk \cite{MarcusWMM11}. CureQL query $q_4$ specifies a time limit of $t$ units, and a maximum number of $k$ rows although, in principle, a crowd query posed can be indefinite aiming to collect rows perpetually.

\section{CrowdCure System Architecture}

\begin{figure*}
\centerline{\includegraphics[height=2in,width=.9\textwidth]{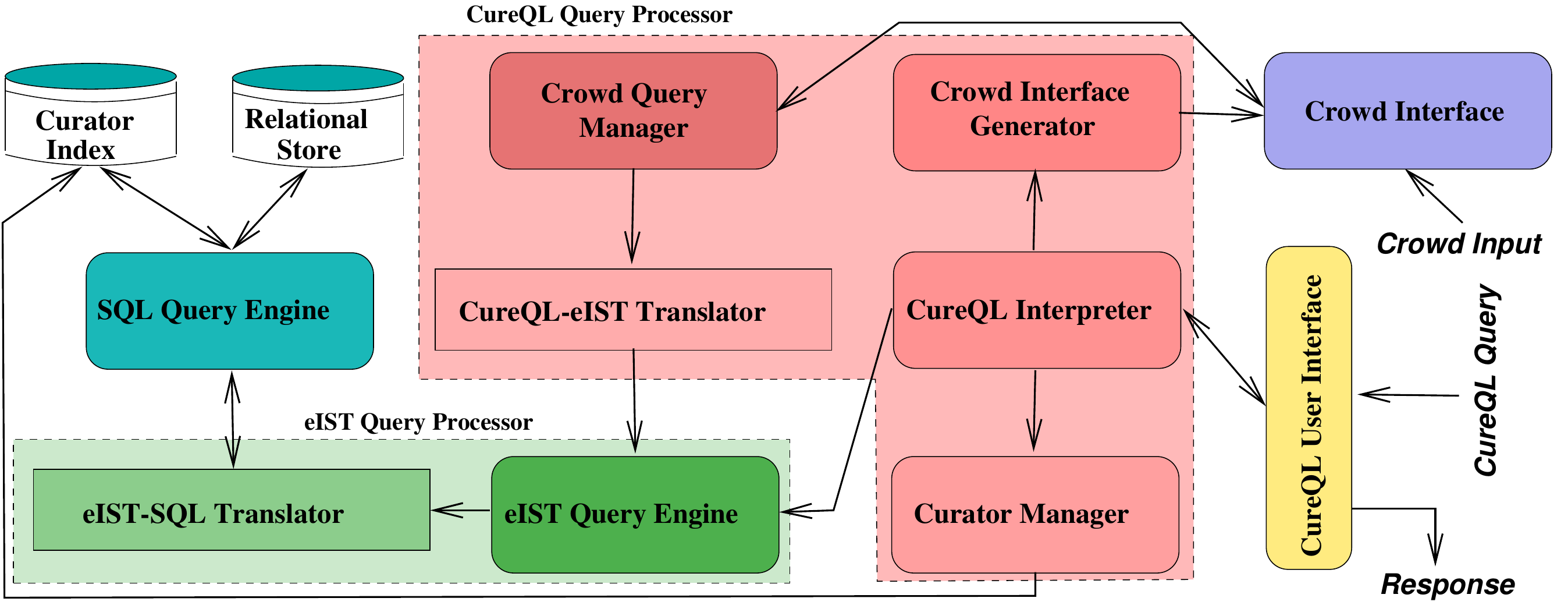}}
\caption{CrowdCure database system architecture.} \label{CCSystem}
\end{figure*}

The overall architecture of CrowdCure system is shown in figure \ref{CCSystem}. In this architecture, we leverage an eIST query processor, which in turn is built as a front-end to a relational query engine, and a relational storage system. Essentially this means that all eIST tables and queries are mapped onto a relational schema for the SQL engine to manage and process. The source vector manipulation and reliability assignment component of a query are managed by the the eIST engine orthogonally by using a separate system index for source information called the {\em Curator Index}. User queries are submitted via the {\em CureQL User Interface} to the {\em eIST Query Engine} transparently of the back-end eIST system after they are preprocessed using the {\em CureQL Interpreter} and translated using the {\em CureQL-eIST Translator}. We briefly describe the {\em CureQL Query Processor} implementation in the next few sections that are unique to CrowdCure system.

\subsection{CureQL Query Processor Implementation}

To process CureQL queries, we implement a minimally invasive stack of two native query processors --  the {\em eIST Query Processor} and the {\em SQL Engine}. To avoid the need for a complete overhaul of eIST, we design an interface between eIST and CureQL, called the {\em CureQL-eIST Translator}, to map all CureQL queries into an equivalent eIST query involving crowd enabled relations. To achieve the effect of the CrowdSQL and Qurk fusion, we have added three subsystems called the {\em Crowd Interface Generator}, {\em Crowd Query Manager} and {\em Curator Manager} to capture the unique ways the {\tt USING ON} and {\tt SOURCE BEFORE} subclauses work. Once a query is received by the {\em CureQL Interpreter}, it is transmitted directly to the eIST engine if no crowd columns are involved in the query scheme. Otherwise, for each table in the {\tt FROM} clause that involves a crowd column, appropriate crowd input forms are created and floated on the web to facilitate curator inputs by the {\em Crowd Interface Generator} unit.

The {\em Crowd Interface Generator} translates {\tt SOURCE} tables into authorization rules and access rights for the curators. For example, if {\em Fred} and {\em Maria} are the curators for a crowd tuple $t$, they will see a request for input in their system while {\em Karen} and {\em Russ} will not. {\tt GROUP BY CLUSTER SOURCE} also works in an analogous way. If {\em Karen} and {\em Russ} are at a higher level in the curator hierarchy, we technically have two choices. In the first case, they will see tuple $t$ once {\em Fred} and {\em Maria} have submitted their input, or their time has expired, and thus allowing {\em Russ} and {\em Karen} to essentially cure the first curation. The other option is to let {\em Karen} and {\em Russ} input their values at anytime and calculate a combined credibility giving higher weight to {\em Russ} and {\em Karen}'s inputs as they are at a strictly higher curation stratum. We adopted the first choice in the current version of CrowdCure system. Thus, depending on the use of the {\tt SOURCE BEFORE} placement in the query, partition information and curator hierarchies are maintained for the entire duration of the query in coordination with the {\em Curator Manager} and {\em Curator Index}. Finally, the {\em Crowd Query Manager} handles the {\tt LIMIT} options along with multi-table queries, {\tt CLUSTER SOURCE}, curator hierarchy, and {\tt SOURCE KEY} consistency enforcement in {\tt SOURCE BEFORE} clause.

\subsection{Query Processing}

Since all tables in CrowdCure are essentially eIST tables (potentially with no {\em Predict} or {\em Archive} components), no queries are processed directly by the {\em SQL Engine}, and so we are able to classify a query into a purely eIST query (with no crowd columns in any tables) or a CureQL query (with having at least one crowd column). Since eventually all queries are translated into eIST queries by the {\em CureQL-eIST Translator} or are directly processed as eIST queries, there is no technical hurdle in preserving the intended semantics of CureQL through staged translation approach adopted. The reliability calculation of the tuples in the query view utilizes the reliability mapping algorithm based on the source credibility, which in turn depends on the past history of curation of the sources as outlined for the basic IST \cite{Sadri91-tkde} and eIST \cite{UncMan-ismis-1994} frameworks, as a final step. For the current version, we have adopted a probabilistic model for reliability calculation, and different or even a parametric model \cite{LakshmananS01b} is equally possible.

\section{Summary}

Our goal in this paper was to introduce a novel hierarchical crowd curation system for fast moving, unreliable and very large PPI data sets. The motivation was to make the predicted data immediately available for use by the end users even before they are decisively cured. Based on multiple prototype implementations and trial runs, we have reasons to believe that this approach delivers more accurate curation with less effort. While limited types of queries are supported in the current prototypes, they supported ad hoc querying in a flexible way through a QBE type graphical interface. A full implementation of CrowdCure system is currently underway. Our ultimate goal is to use STRING \cite{Szklarczyk2015} and DroID \cite{MuraliF2011} as test cases and compare them with CrowdCure in terms of usability, accuracy, scale, curation delay or response time, and expressibility as a future research. This research also opens up a new dimension in crowd computing and novel HIT management interface design that we also plan to pursue.

\bibliographystyle{abbrv}

\begin{thebibliography}{10}

\bibitem{AlonsoMN13a}
O.~Alonso, C.~C. Marshall, and M.~A. Najork.
\newblock A human-centered framework for ensuring reliability on crowdsourced
  labeling tasks.
\newblock In {\em Human Computation and Crowdsourcing: Works in Progress and
  Demonstration Abstracts, An Adjunct to the Proceedings of the First {AAAI}
  Conference on Human Computation and Crowdsourcing, November 7-9, Palm
  Springs, CA, {USA}}, 2013.

\bibitem{AntonyBHP08}
A.~Antony, S.~Basetty, S.~Hartanto, and M.~J. Palakal.
\newblock Computational approach to biological validation of protein-protein
  interactions discovered using literature mining.
\newblock In {\em {ACM} Symposium on Applied Computing (SAC), Fortaleza, Ceara,
  Brazil, March 16-20}, pages 1302--1306, 2008.

\bibitem{AskalidisS2013}
G.~Askalidis and G.~Stoddard.
\newblock A theoretical analysis of crowdsourced content curation.
\newblock In {\em Workshop on Social Computing and User Generated Content},
  2013.

\bibitem{BozzonBCSV13}
A.~Bozzon, M.~Brambilla, S.~Ceri, M.~Silvestri, and G.~Vesci.
\newblock Choosing the right crowd: expert finding in social networks.
\newblock In {\em Joint {EDBT/ICDT} Conferences, {EDBT} Proceedings, Genoa,
  Italy, March 18-22}, pages 637--648, 2013.

\bibitem{CaoXXC15}
D.~Cao, N.~Xiao, Q.~Xu, and A.~F. Chen.
\newblock Rcpi: R/bioconductor package to generate various descriptors of
  proteins, compounds and their interactions.
\newblock {\em Bioinformatics}, 31(2):279--281, 2015.

\bibitem{Chen2004}
H.~Chen and B.~M. Sharp.
\newblock Content-rich biological network constructed by mining pubmed
  abstracts.
\newblock {\em BMC Bioinformatics}, 5(1):1--13, 2004.

\bibitem{DavisWRKLLSJKGHMEM13}
A.~P. Davis, T.~C. Wiegers, P.~M. Roberts, B.~L. King, J.~M. Lay,
  K.~Lennon{-}Hopkins, D.~Sciaky, R.~J. Johnson, H.~Keating, N.~Greene,
  R.~Hernandez, K.~J. McConnell, A.~Enayetallah, and C.~J. Mattingly.
\newblock A ctd-pfizer collaboration: manual curation of 88,000 scientific
  articles text mined for drug-disease and drug-phenotype interactions.
\newblock {\em Database}, 2013, 2013.

\bibitem{FranklinKKRX11}
M.~J. Franklin, D.~Kossmann, T.~Kraska, S.~Ramesh, and R.~Xin.
\newblock Crowddb: answering queries with crowdsourcing.
\newblock In {\em {ACM} {SIGMOD} International Conference on Management of
  Data, Athens, Greece, June 12-16}, pages 61--72, 2011.

\bibitem{HaiderLPLDKLG15}
S.~Haider, Z.~Lipinszki, M.~Przewloka, Y.~Ladak, P.~D'Avino, Y.~Kimata,
  P.~Li{\`{o}}, and D.~Glover.
\newblock {DAPPER:} a data-mining resource for protein-protein interactions.
\newblock {\em BioData Mining}, 8:30, 2015.

\bibitem{JamiesonGSNR12}
D.~G. Jamieson, M.~Gerner, F.~Sarafraz, G.~Nenadic, and D.~L. Robertson.
\newblock Towards semi-automated curation: using text mining to recreate the
  hiv-1, human protein interaction database.
\newblock {\em Database}, 2012, 2012.

\bibitem{UncMan-ismis-1994}
H.~M. Jamil and F.~Sadri.
\newblock Recognizing credible experts in inaccurate databases.
\newblock In {\em ISMIS}, pages 46--55, 1994.

\bibitem{Kalathur08112013}
R.~K.~R. Kalathur, J.~P. Pinto, M.~A. Hernández-Prieto, R.~S. Machado,
  D.~Almeida, G.~Chaurasia, and M.~E. Futschik.
\newblock Unihi 7: an enhanced database for retrieval and interactive analysis
  of human molecular interaction networks.
\newblock {\em NAR}, 2013.

\bibitem{citeulike:3194296}
M.~Krallinger, F.~Leitner, C.~Rodriguez-Penagos, and A.~Valencia.
\newblock Overview of the protein-protein interaction annotation extraction
  task of {BioCreative} {II}.
\newblock {\em Genome biology}, 9 Suppl 2(Suppl 2):S4+, 2008.

\bibitem{KwonKSCW14}
D.~Kwon, S.~Kim, S.~Shin, A.~Chatr{-}aryamontri, and W.~J. Wilbur.
\newblock Assisting manual literature curation for protein-protein interactions
  using bioqrator.
\newblock {\em Database}, 2014, 2014.

\bibitem{LakshmananS01b}
L.~V.~S. Lakshmanan and N.~Shiri.
\newblock A parametric approach to deductive databases with uncertainty.
\newblock {\em IEEE Trans. Knowl. Data Eng.}, 13(4):554--570, 2001.

\bibitem{MarcusWMM11}
A.~Marcus, E.~Wu, S.~Madden, and R.~C. Miller.
\newblock Crowdsourced databases: Query processing with people.
\newblock In {\em Biennial Innovative Data Systems Research Conference,
  Asilomar, CA, USA, January 9-12, Online Proceedings}, pages 211--214, 2011.

\bibitem{McDowall01012009}
M.~D. McDowall, M.~S. Scott, and G.~J. Barton.
\newblock {PIPs}: human protein-protein interaction prediction database.
\newblock {\em NAR}, 37(suppl 1):D651--D656, 2009.

\bibitem{Mehla01052015}
J.~Mehla, J.~H. Caufield, and P.~Uetz.
\newblock Mapping protein-protein interactions using yeast two-hybrid assays.
\newblock {\em Cold Spring Harbor Protocols}, 2015(5), 2015.

\bibitem{MoalJF15}
I.~H. Moal, B.~Jim{\'{e}}nez{-}Garc{\'{\i}}a, and J.~Fern{\'{a}}ndez{-}Recio.
\newblock Ccharppi web server: computational characterization of
  protein-protein interactions from structure.
\newblock {\em Bioinformatics}, 31(1):123--125, 2015.

\bibitem{MuraliF2011}
T.~Murali, S.~Pacifico, J.~Yu, S.~Guest, G.~G. Roberts, and R.~L. Finley.
\newblock {DroID} 2011: a comprehensive, integrated resource for protein,
  transcription factor, {RNA} and gene interactions for drosophila.
\newblock {\em Nucleic Acids Research}, 39(suppl 1):D736--D743, Jan. 2011.

\bibitem{NakatsuI14}
R.~T. Nakatsu and C.~L. Iacovou.
\newblock An investigation of user interface features of crowdsourcing
  applications.
\newblock In {\em 1st International Conference on {HCI} in Business, Held as
  Part of {HCI} International, Crete, Greece, June 22-27}, pages 410--418,
  2014.

\bibitem{Orchardetal14}
S.~E. Orchard, M.~Ammari, B.~Aranda, L.~Breuza, and et~al.
\newblock The {MIntAct} project - {IntAct} as a common curation platform for 11
  molecular interaction databases.
\newblock {\em Nucleic Acids Research}, 42(Database-Issue):358--363, 2014.

\bibitem{ParkW14}
H.~Park and J.~Widom.
\newblock Crowdfill: collecting structured data from the crowd.
\newblock In {\em ACM SIGMOD, Snowbird, UT, USA, June 22-27}, pages 577--588,
  2014.

\bibitem{citeulike:789740}
J.~R. Parrish, K.~D. Gulyas, and R.~L. Finley.
\newblock Yeast two-hybrid contributions to interactome mapping.
\newblock {\em Current opinion in biotechnology}, 17(4):387--393, Aug. 2006.

\bibitem{Peng2014}
W.~Peng, J.~Wang, J.~Cai, L.~Chen, M.~Li, and F.-X. Wu.
\newblock Improving protein function prediction using domain and protein
  complexes in ppi networks.
\newblock {\em BMC Systems Biology}, 8(1):1--13, 2014.

\bibitem{RahmanianD14}
B.~Rahmanian and J.~G. Davis.
\newblock User interface design for crowdsourcing systems.
\newblock In {\em International Working Conference on Advanced Visual
  Interfaces, Como, Italy, May 27-29}, pages 405--408, 2014.

\bibitem{Sadri91-tkde}
F.~Sadri.
\newblock Reliability of answers to queries in relational databases.
\newblock {\em IEEE Trans. Knowl. Data Eng.}, 3(2):245--251, 1991.

\bibitem{Sevimoglu201422}
T.~Sevimoglu and K.~Y. Arga.
\newblock The role of protein interaction networks in systems biomedicine.
\newblock {\em Computational and Structural Biotechnology J.}, 11(18):22 -- 27,
  2014.

\bibitem{SubramaniKMN15}
S.~Subramani, R.~Kalpana, P.~M. Monickaraj, and J.~Natarajan.
\newblock Hpiminer: {A} text mining system for building and visualizing human
  protein interaction networks and pathways.
\newblock {\em Journal of Biomedical Informatics}, 54:121--131, 2015.

\bibitem{citeulike:13903650}
B.~Suter, X.~Zhang, C.~G. Pesce, A.~R. Mendelsohn, S.~P. Dinesh-Kumar, and
  J.-H.~H. Mao.
\newblock {Next-Generation} sequencing for binary {Protein-Protein}
  interactions.
\newblock {\em Frontiers in genetics}, 6, 2015.

\bibitem{Szklarczyk2015}
D.~Szklarczyk, A.~Franceschini, S.~Wyder, K.~Forslund, D.~Heller,
  J.~Huerta-Cepas, M.~Simonovic, A.~Roth, A.~Santos, K.~P. Tsafou, M.~Kuhn,
  P.~Bork, L.~J. Jensen, and C.~von Mering.
\newblock String v10: Protein-protein interaction networks, integrated over the
  tree of life.
\newblock {\em Nucleic Acids Research}, 43(D1):D447--D452, 2015.

\bibitem{citeulike:789741}
J.~Vasilescu and D.~Figeys.
\newblock Mapping protein-protein interactions by mass spectrometry.
\newblock {\em Current Opinion in Biotechnology}, 17(4):394--399, Aug. 2006.

\bibitem{XieHZGFY15}
S.~Xie, Q.~Hu, J.~Zhang, J.~Gao, W.~Fan, and P.~S. Yu.
\newblock Robust crowd bias correction via dual knowledge transfer from
  multiple overlapping sources.
\newblock In {\em {IEEE} Intl. Conf. on Big Data, CA, USA, October 29 -
  November 1}, pages 815--820, 2015.

\bibitem{ZhangLYW15}
Y.~Zhang, H.~Lin, Z.~Yang, and J.~Wang.
\newblock Integrating experimental and literature protein-protein interaction
  data for protein complex prediction.
\newblock {\em {BMC} Genomics}, 16({S-2}):S4, 2015.

\end{thebibliography}

\end{document}